\documentclass[12pt]{iopart}

\usepackage{iopams}  
\usepackage{cite}
\usepackage{graphicx}

\begin{document}

\title[Sensitivity of a heliotron configuration to ferritic-steel perturbations]
{Sensitivity of a low-shear heliotron configuration to localised ferritic-steel perturbations}

\author{A. Matsuyama$^{1}$, F. Tanji$^{1}$, Y. Nakamura$^{1}$, S. Inagaki$^{2}$, Yusuke Yamashita$^{1}$, Akihisa Yamamoto$^{1}$, S. Kobayashi$^{2}$, F. Kin$^{2}$, S. Kado$^{2}$, S. I. Inagaki$^{2}$, S. Konoshima$^{2}$, T. Mizuuchi$^{2}$, K. Nagasaki$^{2}$}

\address{$^{1}$Graduate School of Energy Science, Kyoto University, Gokasho, Uji 611-0011, Japan}
\address{$^{2}$Institute of Advanced Energy, Kyoto University, Gokasho, Uji 611-0011, Japan}
\ead{akinobu.matsuyama@gmail.com}

%%%%%% ABSTRACT
\begin{abstract}
The influence of ferritic steel on low-shear stellarator/heliotron magnetic configurations is investigated for the Heliotron J device using a point dipole magnetisation model. By numerically evaluating ferritic steel plates assumed at several locations inside the Heliotron J vacuum vessel, the changes in the rotational transform and magnetic island width are shown to be sensitive to the installation location. This location sensitivity arises from the toroidal variation of poloidal mode coupling between the background nonaxisymmetric field and ferritic-steel perturbation, rather than being determined solely by the perturbation amplitude. The resulting mode coupling can enhance the resonant vacuum magnetic perturbation at specific locations. Ferritic steel plates placed on the outer side of a straight section produce the most significant changes in the magnetic topology and exhibit the highest sensitivity to violations of the $M=4$ toroidal periodicity. Additionally, we show that appropriate arrangements of passive magnetic dipoles can reduce the effective helical ripple while preserving the vacuum magnetic well depth in Heliotron J, and can induce a stellarator-asymmetric boundary perturbation in low-field experiments.
\end{abstract}

\submitto{Plasma Phys. Control. Fusion}
\maketitle

\section{Introduction}
Structural materials for in-vessel components must satisfy the demanding engineering and nuclear requirements of future fusion reactors. In addition to resistance to neutron irradiation, high-temperature strength, and weldability, they must exhibit low-activation characteristics \cite{Kohyama1996,Tanigawa2011} to facilitate reactor maintenance and radioactive waste management. Reduced-activation ferritic/martensitic (RAFM) steels \cite{Tanigawa2017,Tavassoli2014}, such as F82H \cite{Jitsukawa2002} and EUROFER-97 \cite{Mergia2008}, are among the leading candidate materials for blanket systems and other in-vessel components \cite{Gorley2021,Giancarli2012}.

For the practical application of RAFM steels, it is essential to quantify the influence of ferritic magnetisation on plasma confinement. In present experimental devices, magnetic materials are primarily treated as sources of error fields. However, tokamak experiments have suggested that magnetic materials can provide an additional degree of freedom for magnetic-field control \cite{Turner1978,Sheffield1993,Ane1994,Tobita2003}. In JFT-2M \cite{Sato1998,Kawashima2001} and JT-60U \cite{Shinohara2007}, for example, ferritic steel tiles were installed to reduce the toroidal field ripple, which was shown experimentally and numerically to affect energetic-particle confinement \cite{Shinohara2003,Shinohara2007} and plasma rotation \cite{Yoshida2006,Honda2014}. These studies also demonstrated that normal plasma operation is possible with ferritic steel, provided that appropriate wall conditioning procedures are applied \cite{Tsuzuki2006}.

The use of RAFM steels is also anticipated in stellarator/heliotron reactors \cite{Harmeyer1999,Ji2017,Landreman2026}. Because stellarators generate rotational transform using external coils and possess intrinsically three-dimensional magnetic field structures, the influence of magnetic materials is nontrivial. Previous studies have indicated that magnetic perturbations from ferritic materials can lead to the formation of sizable magnetic islands \cite{Harmeyer1999}. From an engineering perspective, the limited space and complex geometry inside the vacuum vessel require consideration of experimental constraints in the design of in-vessel components. These considerations motivate the present investigation of ferritic-steel effects in an operating stellarator/heliotron device, providing physical insight relevant to future reactor design.

In this study, the Heliotron J device \cite{Wakatani2000,Obiki2001} is used to investigate the influence of magnetic materials on stellarator/heliotron configurations. Heliotron J is a helical-axis heliotron device operating at Kyoto University, which is characterised by low magnetic shear and a vacuum magnetic well. These characteristics are shared with modular quasi-isodynamic stellarators, including W7-X \cite{Beidler2021}. We numerically evaluate the effects of ferritic-steel perturbations using a point dipole magnetisation model, focusing on their two complementary aspects: first, as a source of unintended error fields that can break magnetic surfaces, and second, as a beneficial passive element for tailoring the magnetic configuration.

In tokamaks, error fields are known to lock magnetohydrodynamic (MHD) modes to a specific phase and can ultimately be a cause of plasma disruptions \cite{Bandyopadhyay2025}. In stellarators, the impact on confinement is more direct, as these perturbations can break magnetic surfaces of the vacuum equilibrium configuration. The requirements for coil installation accuracy during the design stage \cite{Yamazaki1993,Andreeva2009,Shoji2023} and the precision of the actual configuration during operation \cite{Jaenicke1993,Morisaki2010,Pedersen2016} have been experimentally investigated. (See \cite{Lazerson2018} and references therein.) The correction of error fields by three-dimensional external coils and associated plasma response have also been an important focus in tokamaks \cite{Boozer2011,Pharr2024}. While the response of the configuration to error fields is non-uniform in the poloidal direction in tokamaks \cite{Boozer2011}, it becomes non-uniform in both the poloidal and toroidal directions in stellarators \cite{Zhu2019,Cao2026}. The installation of ferritic steel in the Heliotron J device shows similar non-uniformity such that the magnetic response depends more significantly on the installation location than on the magnetic perturbation amplitude. 

Another aspect of ferritic materials is their potential for passive configuration control. In particular, in present-day experiments operated at relatively low magnetic fields (e.g., around 0.4 T), the magnetisation of ferritic steel, such as F82H, is close to saturation, whereas the field produced by ferritic steel can become comparable to the equilibrium field. Under such specific conditions, magnetic materials can provide an additional degree of freedom for tailoring three-dimensional magnetic configurations. {Recent studies have explored} localised magnetic elements such as permanent magnets \cite{Helander2020,Hammond2020,Qian2022}, non-plasma-surrounding coils\cite{Gates2025,Kruger2025}, and passive superconductors \cite{Kaptangolu2025}{,} for constructing and improving stellarator configurations\cite{Ku2009,Elder2024,Boozer2024}{.} {Motivated by these developments, we} also examine the possibility of improving the effective helical ripple and deforming the plasma boundary shape in Heliotron J using ferritic{-steel} materials.  

The remainder of this paper is organised as follows. Section \ref{sec:method} introduces the numerical method used to calculate the magnetic fields produced by ferritic steel plates. Section \ref{sec:result} investigates the influence of ferritic steel plates as sources of error fields in Heliotron J. Section \ref{sec:application} illustrates the potential of ferritic steel as an element for the passive modification of the Heliotron J configuration. Finally, section \ref{sec:conclusion} summarizes the conclusions of this study.

\section{Simulation methods}\label{sec:method}
\subsection{Heliotron J configuration}
Heliotron J \cite{Wakatani2000,Obiki2001} is a heliotron device with a helical magnetic axis, major radius of $R_0=1.2$ m, and average minor radius of $a=0.13$ m. The coil system consists of a continuous helical field coil with poloidal and toroidal period numbers of $L=1$ and $M=4$, respectively, complemented by three pairs of vertical field coils and 16 toroidal field coils of two types (TA and TB coils). While Heliotron J is typically operated at $B_0=1.25$ T for electron cyclotron heating, operation at lower field strengths is also possible.

 In this study, we consider magnetic configurations with $B_0$ down to 0.42 T to examine regimes where the ferritic magnetic field constitutes a larger fraction of the equilibrium magnetic field. Lower-field operation, e.g., at $B_0\sim0.1$ T, is not quantitatively evaluated in the present study because the ferritic steel enters the unsaturated magnetisation regime, where a detailed treatment of the magnetisation becomes important.
 
Heliotron J was designed \cite{Wakatani2000} with a particular emphasis on the formation of a vacuum magnetic well through the selection of the helical coil modulation. As a consequence, toroidicity on the order of the geometrical inverse aspect ratio remains, which can limit energetic-particle confinement and neoclassical transport. Nevertheless, Heliotron J shares several key characteristics with modern quasi-isodynamic stellarators \cite{Goodman2023}, including low magnetic shear, a vacuum magnetic well, and the use of a toroidal mirror component. The TA and TB coils generate a toroidal mirror field with $(m,n)=(0,4)$ as a sideband to the helical field \cite{Wakatani2000}, where {$m$ and $n$ are the poloidal and toroidal} mode numbers, respectively. This sideband field aligns the minima of the magnetic-field ripple, $B_{\rm min}$, along the magnetic field lines in a manner similar to $\sigma$-optimisation \cite{Mynick1982}. The configuration with a TA:TB current ratio of 5:2 is referred to as the ``standard configuration," while the TA:TB = 5:1 configuration enhances the toroidal mirror component and modifies the trapped-particle confinement properties \cite{Yokoyama2000}. Unless otherwise noted, the analyses in section \ref{sec:result} focus on the standard configuration. The TA:TB = 5:1 (``high-bumpiness") configuration will be considered in section \ref{sec:application}.

In this study, the vacuum magnetic field generated by the coil system of Heliotron J was calculated using {\small KMAG} \cite{Nakamura1992}. In {\small KMAG}, coils are discretised based on a finite-cross-section model \cite{Todoroki1987}, and the magnetic field is calculated by numerical integration based on the Biot-Savart law. The {\small KMAG} code searches for the magnetic axis and the last closed magnetic surface while evaluating the rotational transform profile and Poincar\'{e} plots. The vacuum field calculated by {\small KMAG} is used as the external magnetic field, ${\bf B}_{\rm ext}$, and the magnetic response to the ferritic steel installation is analysed by superposing the ferritic field onto this external field. Where required, the vacuum magnetic field calculated by {\small KMAG} is further transformed into Boozer coordinates using {\small VMEC} \cite{Hirshman1983} and {\small BOOZXFORM} \cite{Sanchez2000}.

\subsection{Point dipole magnetisation model}
In this study, we employ a magnetic dipole approximation to evaluate the ferritic-steel effects on the magnetic configurations. The point-dipole representation provides a computationally efficient means of evaluating the magnetic perturbation produced by distributed ferritic components. For the present calculation, the magnetisation {assigned to each point dipole} is determined solely by the local external magnetic field, ${\bf B}^{\rm ext}$. {It should be noted that in general, the magnetisation distribution in magnetic materials should be determined self-consistently by accounting for both the mutual interactions among magnetic bodies and the magnetic field produced by the induced magnetisation itself.}
 
With the magnetisation prescribed in this manner, the induced magnetisation ${\bf M}({\bf r}^\prime)$ in a ferritic element located at position ${\bf r}^\prime$ is expressed as
\begin{equation}
 {\bf M}({\bf r}^\prime) = M_s f(|{\bf B}^{\rm ext}({\bf r}^\prime)|) {{\bf B}^{\rm ext}({\bf r}^\prime) \over |{\bf B}^{\rm ext}({\bf r}^\prime)|}, \label{eq:magnet}
\end{equation}
where {$M_s$ is the saturation magnetisation in A/m. In the present calculation, we use $\mu_0$}$M_s=1.976$ T \cite{Urata2003}. The function $f(|{\bf B}^{\rm ext}|)$ represents the magnetisation curve of the ferritic steel. Neglecting hysteresis, it can be modelled using a function such as $f(|{\bf B}^{\rm ext}|)=\tanh(|{\bf B}^{\rm ext}|/B_{\rm sat})$. For F82H steel, the magnetisation reaches saturation at external magnetic fields of approximately 0.25--0.3 T{; this behaviour is represented using} $B_{\rm sat} \simeq 0.1$ T \cite{Shiba1997,Nakayama1999}. {For the lowest-field case considered in this study ($B_0=0.42$ T), the minimum external field at the ferritic-steel locations is approximately 0.27 T. Even at this minimum value, the present magnetisation model gives $M/M_s=\tanh(0.27/0.1)\simeq0.991$. Thus, the ferritic steel is already close to magnetic saturation throughout the parameter range considered here, and the calculated magnetisation is insensitive to moderate variations in $B_{\rm sat}$.}

In our model, a ferritic steel plate is represented as an assembly of point magnetic dipoles. The ferritic plates are discretised into cubic elements with a characteristic size of approximately 1 cm. The magnetisation vector ${\bf M}({\bf r}^\prime)$ of each element at its centre ${\bf r}^\prime=(x^\prime,y^\prime,z^\prime)$ is evaluated using Eq.~(\ref{eq:magnet}). The magnetic field ${\bf B}_m({\bf x})$ generated at an observation point ${\bf x}$ is then obtained by superposing the magnetic dipole fields of all elements:
\begin{equation}
{\bf B}_m ({\bf x}) = {{\mu_0} \over 4\pi} \int_{V^\prime} \left[ {3({\bf M} \cdot {\bf R} ){\bf R} \over R^5}  - {{\bf M} \over R^3} \right] dV^\prime, \label{eq:Bferrite}
\end{equation}
where ${\bf R}={\bf x}-{\bf r}^\prime$ and $R=|{\bf R}|$. The volume integral, $\int dV^\prime = dx^\prime dy^\prime dz^\prime$, is numerically evaluated as a summation over the discretised dipole elements.

\subsection{Benchmark with the FEMAG code {and limitations of the point-dipole model}}
The {\small KFERRITE} code was developed to implement Eqs.~(\ref{eq:magnet}) and (\ref{eq:Bferrite}) and interface with the {\small KMAG} code. To verify its implementation, we reproduce the magnetic field configuration of the JT-60U ferritic steel insertion experiment for the E45119 equilibrium and compare our results with those obtained using {\small FEMAG} \cite{Urata2003,Honda2014}. {The toroidal magnetic field is $B_T = 2.6$ T.}

\begin{figure}[htbp]
\begin{center}
\includegraphics[clip,width=10.0cm]{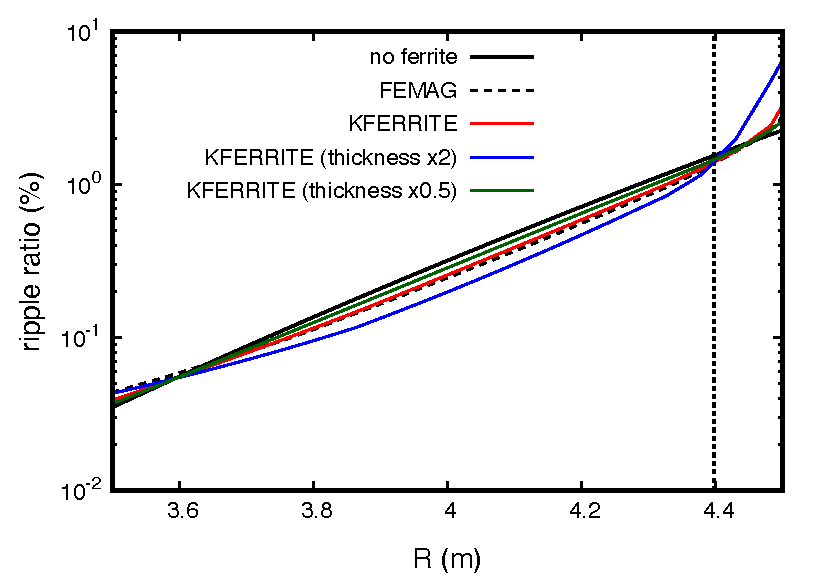}
\caption{Comparison of toroidal field ripple calculations for the JT-60U E45119 equilibrium \cite{Honda2014} obtained using KFERRITE (red) and FEMAG (dashed black curve). Additional KFERRITE results are shown for the doubled (blue) and halved (green) ferritic steel tile thicknesses. The vertical line corresponds to the position of the plasma surface.}
\label{fig:JT-60U_ripple}
\end{center}
\end{figure}

The benchmark results are presented in figure~\ref{fig:JT-60U_ripple}, which shows the toroidal field ripple on the equatorial plane ($Z=0$) as a function of the major radius $R$. Here, the ripple ratio is defined as $\delta_{\rm ripple}=(B_{\rm max}-B_{\rm min})/(B_{\rm max}+B_{\rm min})$, where $B_{\rm max}$ and $B_{\rm min}$ denote the maximum and minimum magnetic field strengths in the toroidal direction, respectively. The {\small KFERRITE} results show quantitatively good agreement with the FEMAG calculations. 

A sensitivity analysis was also conducted by varying the thickness of the ferritic steel tile to double and half of its original value. The results indicate that the ripple amplitude scales approximately in proportion to the tile thickness. This behaviour is consistent with what is described in the {\small FEMAG} code documentation \cite{Urata2003}. Physically, because the toroidal magnetic field is applied predominantly tangentially to the ferritic tile surface, the resulting magnetic poles are formed near the tile edges, whose strength is proportional to the area of the poles. This implies that the induced magnetic perturbation is proportional to the volume of the material near these edges and thus to the tile thickness.

While the above benchmark supports the applicability of the point dipole model in the strong external-field regime, where the applied magnetic field is comparable to or larger than the characteristic field produced by the magnetised material ($|{\bf B}|\gtrsim \mu_0 M_s$), a self-consistent evaluation of the magnetisation distribution becomes important at lower magnetic fields. To estimate the magnitude of this effect, we additionally performed a simplified self-consistent magnetostatic finite-element method (FEM) calculation for a rectangular ferritic-steel plate with dimensions representative of those considered in this study. The magnetisation distribution was iteratively updated from the local magnetic field obtained from the FEM solution using the same phenomenological magnetisation relation as in Eq.~(\ref{eq:magnet}), thereby including the self-demagnetising field within the plate.

For an applied field of 0.27 T parallel to the plate surface, corresponding to the weakest external field considered in the present Heliotron J calculations, the self-consistent total magnetic moment was approximately 13 $\%$ smaller than that obtained with the prescribed magnetisation approximation. At a distance of 10 cm from the plate surface, the magnitude of the ferritic magnetic field was reduced from approximately 10.4 mT to 9.3 mT, corresponding to a difference of approximately 11 $\%$. These results confirm that the predominantly in-plane magnetisation expected for the conformally installed plates in Heliotron J substantially mitigates, although does not eliminate, the self-demagnetising effect, consistent with the geometrical dependence of the demagnetising factor for a thin plate \cite{Aharoni1998}. The magnitude of the correction under these conditions supports the use of the point dipole model as a first-order quantitative model for the purpose of the present study, while more detailed self-consistent modelling would be required for accurate prediction of the ferritic-steel perturbation. At an applied field of 1.25 T, by contrast, the same self-consistent FEM calculation yielded corrections to the total magnetic moment below 0.2 $\%$ for both in-plane and normal magnetisation, confirming that self-demagnetisation effects are negligible for the $B_0=1.25$ T cases considered in the present calculations. 

\section{Magnetic field response to ferritic steel plates}\label{sec:result}
\subsection{Ferritic steel plate model in Heliotron J}
\begin{figure}[htbp]
\begin{center}
\includegraphics[clip,width=10.0cm]{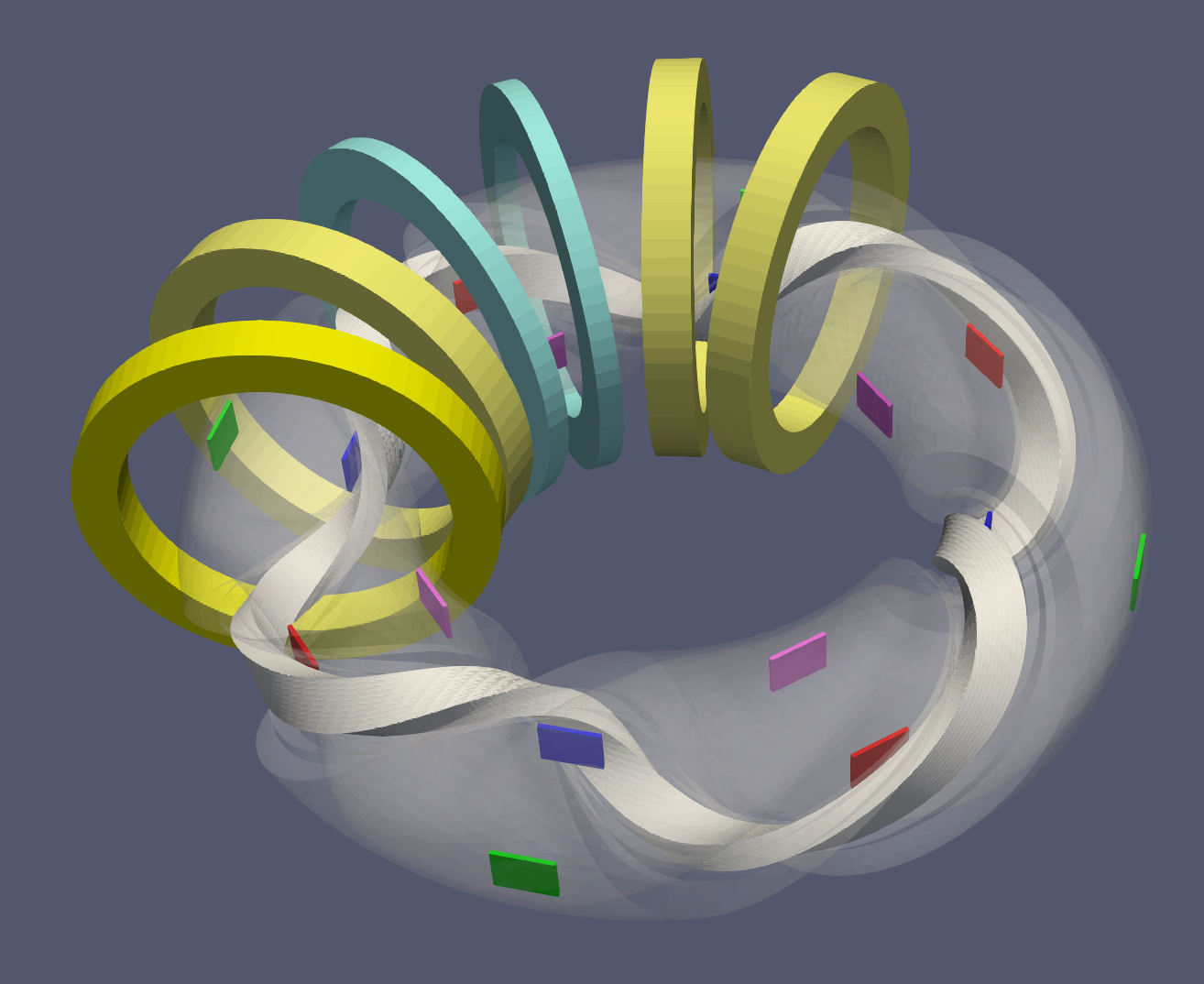}
\caption{Installation locations of the test ferritic steel plates. Four plates are placed at toroidally periodic positions: (i) inner side of the corner section (C-I; blue), (ii) outer side of the corner section (C-O; green), (iii) inner side of the straight section (S-I; magenta), and (iv) outer side of the straight section (S-O; red). The $L=1$ helical coil (gray), two pairs of the TA coils (yellow), and a pair of the TB coils (light blue) are also displayed. }
\label{fig:ferrite_stl}
\end{center}
\end{figure}

In this section, ferritic steel plates are introduced numerically inside the Heliotron J vacuum vessel to examine their influence on the standard configuration (TA:TB = 5:2). Test plates with a width of 20 cm, height of 15 cm, and thickness of 15 mm are placed at the locations shown in figure~\ref{fig:ferrite_stl}. From a top-down perspective, Heliotron J exhibits a square-shaped toroidal structure with $M=4$ periodicity. The helical field coil is positioned on the outer side of the torus in the straight sections and on the inner side in the corner sections of the torus. Based on these geometrical characteristics, four representative locations are considered: (i) the inner side of a corner section (C-I; blue), (ii) the outer side of a corner section (C-O; green), (iii) the inner side of a straight section (S-I; magenta), and (iv) the outer side of a straight section (S-O; red). Note that the external magnetic field strength is the lowest at the outer side of the corner sections. {Unless otherwise noted, ferritic-steel plates are placed at four toroidally periodic positions for each case.}

The dimensions of the ferritic plates used in this study are selected with reference to the JFT-2M tokamak, which has a device scale ($R=1.31$ m, $a=0.3$ m, and $B_t=2.2$ T) relatively close to that of Heliotron J. In the JFT-2M experiments\cite{Shinohara2003,Tsuzuki2006}, ferritic steel tiles with thicknesses of 6, 8, and 10.5 mm were installed over nearly the entire inner wall of the vacuum vessel. The ferritic plates considered in our model are somewhat thicker than the thickest tiles used in JFT-2M in order to produce sufficiently large magnetic perturbations for the present sensitivity study. {In section \ref{sec:application}, relatively large plate thicknesses up to 30-40 mm are also considered to explore the achievable range of magnetic perturbations. These large thicknesses are introduced only for the present numerical study  and do not represent a directly installable configuration in Heliotron J.}

\begin{figure}[htbp]
\begin{center}
\includegraphics[clip,width=15.0cm]{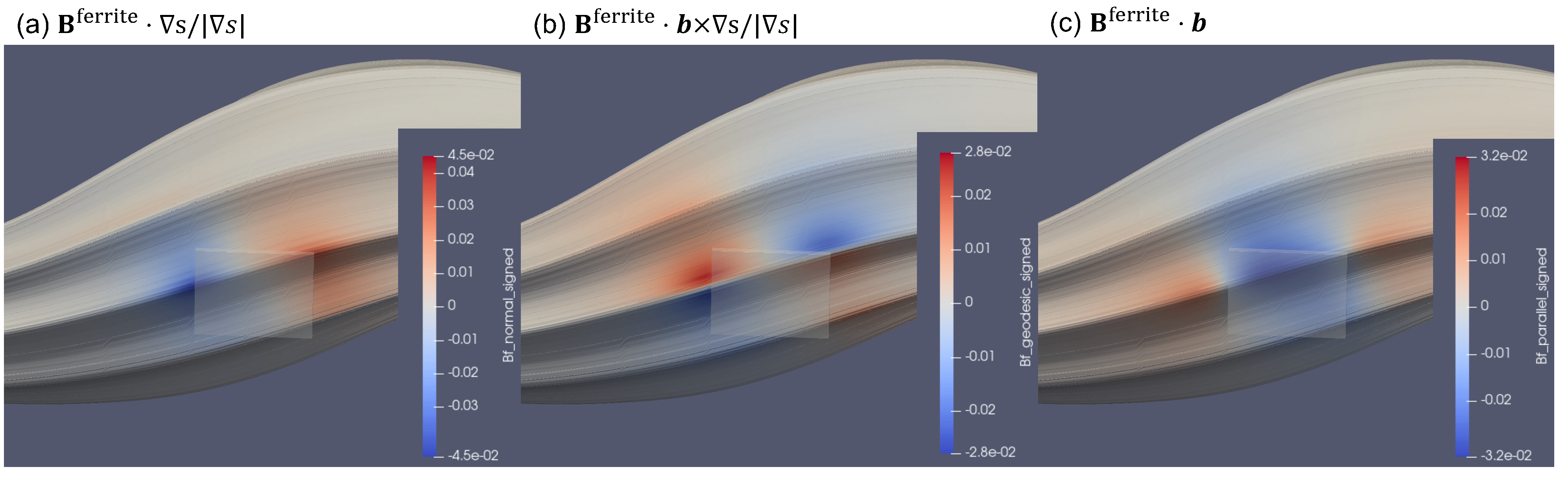}
\caption{Three components of the magnetic field generated by a ferritic steel plate installed on the inner side of a straight section, evaluated on the unperturbed plasma boundary surface at $B_0=1.25$ T: (a) normal, (b) geodesic, and (c) parallel components. Note that the toroidal magnetic field generated by the coil system is directed in the negative toroidal direction ($B_\varphi < 0$). The colour scale, however, represents ${\bf B}_{\rm ferrite}\cdot {\bf b}$; thus, negative values indicate a ferritic perturbation antiparallel to the local equilibrium field. The ferritic steel plates are displayed in a translucent colour.}
\label{fig:ferrite_SI_single}
\end{center}
\end{figure}

\subsection{Spatial structure of localised ferritic-steel perturbation}
Figure~\ref{fig:ferrite_SI_single} illustrates the magnetic field generated by the ferritic plate installed on the inner side of the straight section at $B_0=1.25$ T. Here the ferritic field components are displayed as contour plots projected onto the three-dimensional surface representing the plasma boundary, which is determined for the unperturbed vacuum magnetic field configuration in the absence of ferritic steel plates. Although the ferritic plate is modelled in {\small KFERRITE} as an assembly of numerous saturated dipoles, the resulting magnetic field exhibits the characteristic structure of a macroscopic dipole field {with a characteristic scale comparable to the size of} the ferritic plate. Opposite magnetic poles are formed near the two ends of the plate, and the dipole axis is slightly tilted from the geometrical toroidal direction, reflecting the local pitch of the helical magnetic field. 

Panels (a)--(c) show contour plots of the ferritic magnetic field projected onto the surface-normal, geodesic, and field-aligned (parallel) directions on the plasma surface, respectively. While the amplitudes of all three components remain of the same order, the radial (normal) component is found to be the strongest ($\delta B\simeq 0.05$ T). The radial perturbation ${\bf B}^{\rm ferrite}\cdot \nabla s$ appears as a pair of positive and negative lobes associated with the two magnetic poles of the ferritic plate, a structure expected to produce a localised wavy distortion as field lines pass over the plate. A similar positive-negative pair structure is observed in the geodesic component. This symmetry suggests that these components produce localised distortions associated with a magnetic dipole.

The parallel component exhibits a different behaviour in figure~\ref{fig:ferrite_SI_single}(c). Over most of the area facing the ferritic plate, the ferritic magnetic field is directed opposite to the local equilibrium magnetic field, indicating that the ferritic plate weakens the parallel magnetic field. Although regions where the ferritic field strengthens the equilibrium field exist near the magnetic poles, their peak amplitude is about half of that in the antiparallel region near the centre of the plate ($\delta B\sim 0.03$ T). {This reduction is consistent with the tendency of the magnetised plate to draw magnetic flux toward itself.} As shown below, {the resulting magnetic} perturbations {displace} the magnetic surfaces toward the ferritic plates.

\subsection{Influence on the rotational transform}
\begin{figure}[htbp]
\begin{center}
\includegraphics[clip,width=15.0cm]{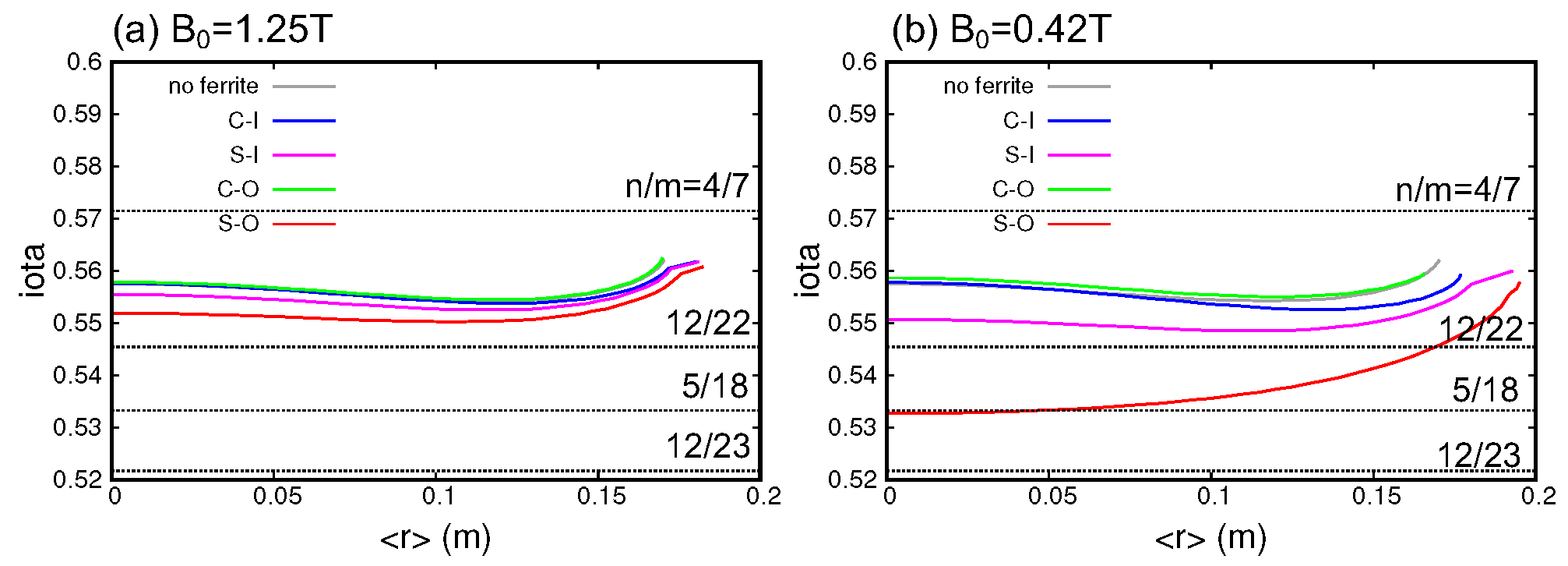}
\caption{Rotational transform profiles for ferritic steel plates installed at different locations: (a) $B_0=1.25$ T and (b) $B_0=0.42$ T. The line colors correspond to the ferritic plate locations according to the convention used in figure \ref{fig:ferrite_stl}. {The} reference curve (gray){, obtained} without ferritic plates{, nearly} overlaps the C-O curve.}
\label{fig:ferrite_iota}
\end{center}
\end{figure}

In low-magnetic-shear devices, such as Heliotron J, the magnetic configuration is intrinsically sensitive to even subtle modifications of the rotational transform ($\iota$) profile. Figure~\ref{fig:ferrite_iota} presents the rotational transform profiles for the four representative ferritic plate locations at magnetic-axis field strengths of $B_0=1.25$ T and 0.42 T.

\begin{figure}[htbp]
\begin{center}
\includegraphics[clip,width=15.0cm]{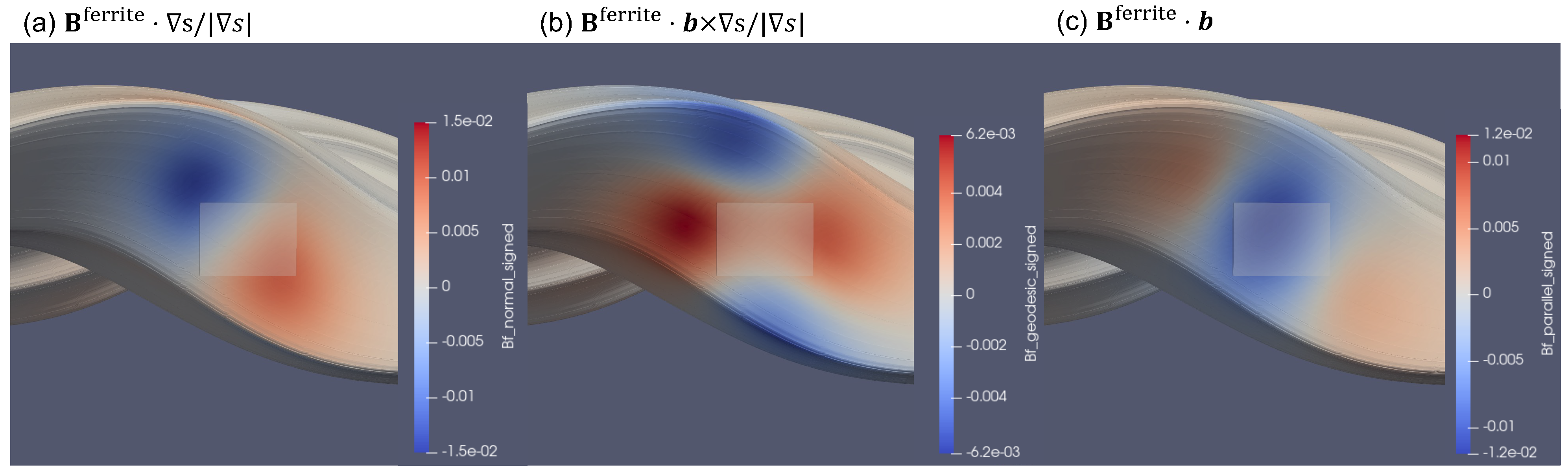}
\caption{Three components of the magnetic field generated by a ferritic steel plate installed on the outer side of a straight section (S-O), evaluated on the unperturbed plasma boundary surface. See also figure \ref{fig:ferrite_SI_single}.}
\label{fig:ferrite_SO_single}
\end{center}
\end{figure}

The results show a dependence on the plate location. The most pronounced modification occurs when the ferritic plate is positioned on the outer side of the straight section (S-O). For instance, at $B_0=0.42$ T, this case induces a change in the rotational transform of approximately $\Delta \iota \sim 0.03$ on the magnetic axis. In contrast, the modification is small when the plate is placed on the outer side of the corner section (C-O). Notably, the magnitude of $\Delta \iota$ does not scale with the strength of the ferritic magnetic field itself. On the plasma surface, the {maximum} magnetic field strengths produced by the ferritic plate are 0.11 T for C-I, 0.068 T for C-O, 0.045 T for S-I, and only 0.0145 T for S-O. No correlation is found between the change in the rotational transform and the perturbation strength at the plasma surface, regardless of which of the three magnetic field components is examined. These results indicate that the magnetic response {cannot be characterised by the magnetic field strength produced by the ferritic steel and suggest a role of the spatial structure of the magnetic perturbation relative to the three-dimensional equilibrium field.} Figure \ref{fig:ferrite_SO_single} {shows the corresponding three magnetic-field components for the S-O case.}

\subsection{Vacuum magnetic islands induced by ferritic-steel perturbations}
\begin{figure}[htbp]
\begin{center}
\includegraphics[clip,width=15.0cm]{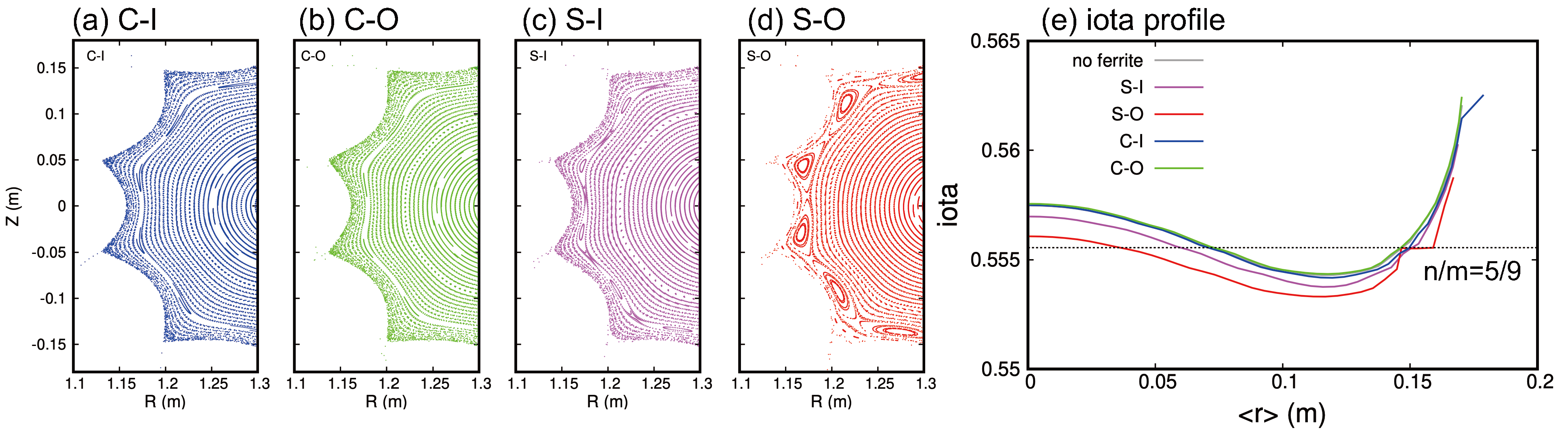}
\caption{Changes in the magnetic configuration at $B_0=1.25$ T when a ferritic steel plate is installed at only one toroidal location: (a)--(d) Poincar\'{e} plots of magnetic field lines and (e) rotational transform profiles.}
\label{fig:ferrite_island}
\end{center}
\end{figure}

\begin{figure}[htbp]
\begin{center}
\includegraphics[clip,width=15.0cm]{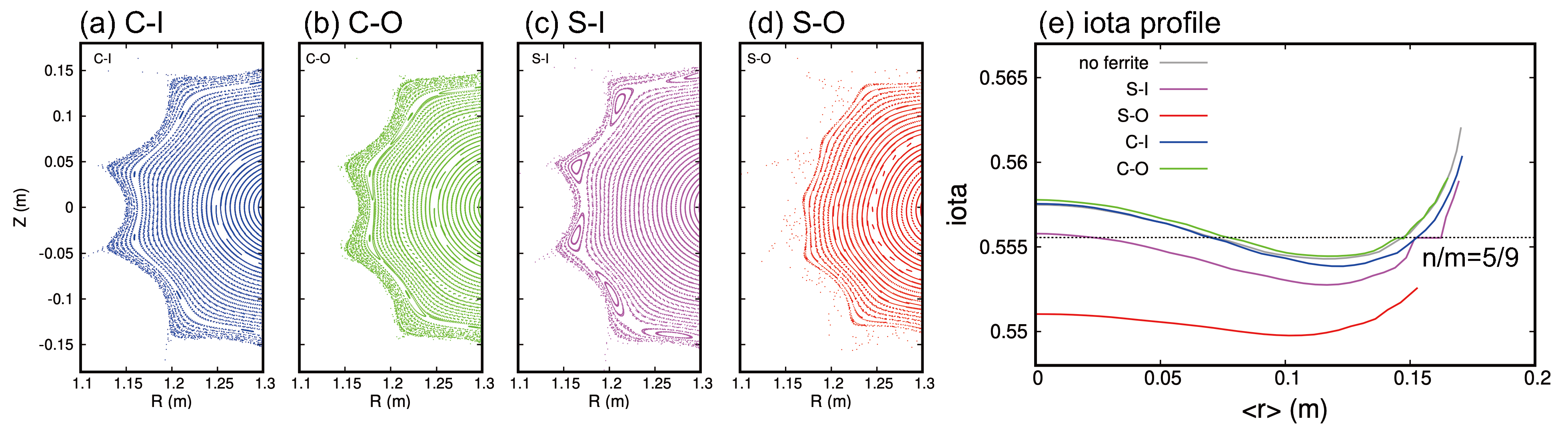}
\caption{Changes in the magnetic configuration at $B_0=0.42$ T when a ferritic steel plate is installed at only one toroidal location: (a)--(d) Poincar\'{e} plots of magnetic field lines and (e) rotational transform profiles.}
\label{fig:ferrite_island_third}
\end{center}
\end{figure}

Figures~\ref{fig:ferrite_island} and \ref{fig:ferrite_island_third} show the Poincar\'{e} plots and rotational transform profiles for configurations in which a ferritic plate is installed at only one toroidal location at $B_0=1.25$ and 0.42 T, respectively. {For the Poincar\'{e} plots, magnetic field lines were traced for 1000 toroidal turns. The initial points were placed at 5 mm intervals along the minor-radial direction from the magnetic axis on a poloidal cross-section in the straight section.}

Because the plate is {localised in the toroidal direction and does not preserve} the intrinsic $M=4$ periodicity of Heliotron J, the ferritic perturbation {contains} a broad spectrum of toroidal Fourier components{, including $n=5$, which can resonate with the $\iota=n/m=5/9$ rational surface}. 

A strong dependence on the plate location is again observed. Although the rotational transform profile {crosses} the $(m,n) =(9,5)$ resonance in all cases, a large magnetic island is formed only when the plate is located on the outer side of the straight section (S-O), as seen in figure~\ref{fig:ferrite_island}(d). In the case of $B_0 = 0.42$ T, the outer magnetic surfaces are broken and the effective confinement volume {also} shrinks [figure \ref{fig:ferrite_island_third}(d)]{.} To quantify {the perturbation}, the Fourier components of ${\bf B}^{\rm ferrite}\cdot\nabla s/B_0$ were {first} evaluated. {The overall} amplitudes of the $n=1$ and $n=5$ components are of the order of $10^{-3}$ and show no strong dependence on the plate location. {The strong variation in the island width therefore cannot be explained by the overall amplitude of the $n=5$ perturbation alone.}  

\begin{figure}[htbp]
\begin{center}
\includegraphics[clip,width=15.0cm]{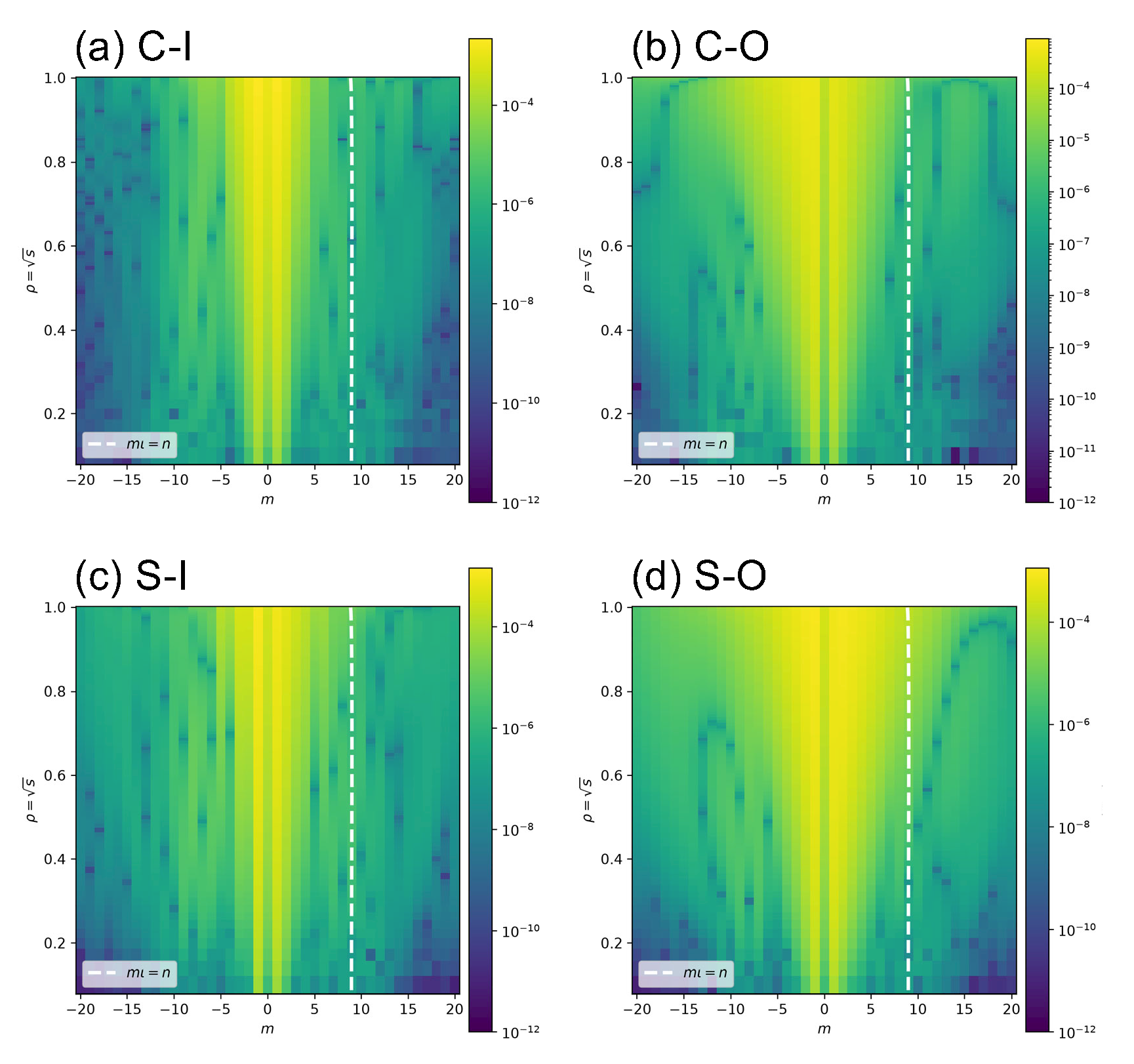}
\caption{{Poloidal mode spectra of the $n=5$ ferritic magnetic perturbation for (a) C-I, (b) C-O, (c) S-I, and (d) S-O. Contours show ${\bf B}^{\rm ferrite}\cdot\nabla s/B_0$ as functions of the poloidal mode number $m$ and $\rho$, defined as the square root of the normalised toroidal flux. The dashed line indicates the resonance condition $m\iota-n=0$ for $n=5$.}}
\label{fig:resonance}
\end{center}
\end{figure}

To further clarify the origin of the strong location dependence, we examine the poloidal Fourier spectrum of the $n=5$ ferritic-steel perturbation. Since a single ferritic plate is localised in the toroidal direction, its perturbation has a broad toroidal mode spectrum for all four installation locations. However, comparison of the toroidal spectra did not reveal a significant location dependence that could account for the markedly different magnetic-island responses. We therefore focus on the poloidal mode structure of the $n=5$ component. Figure \ref{fig:resonance} shows contours of ${\bf B}^{\rm ferrite}\cdot\nabla s/B_0$ as functions of the poloidal mode number $m$ and $\rho$, defined as the square root of normalised toroidal flux. The C-I and S-I cases exhibit relatively narrow poloidal spectra, whereas the C-O perturbation spreads mainly toward negative $m$. In contrast, the S-O case exhibits a broad poloidal spectrum, including a positive-$m$ component. Consequently, the resonant $(m,n)=(9,5)$ component at the $\iota=n/m=5/9$ rational surface is enhanced in the S-O case. The ordering of the resonant $(9,5)$ amplitudes among the four locations is consistent with the corresponding magnetic-island widths obtained from the Poincar\'{e} plots. This indicates that the strong location dependence is associated with the toroidal variation of the poloidal mode coupling of the ferritic perturbation to the three-dimensional background magnetic field, rather than with the overall amplitude of the $n=5$ perturbation. Numerical sensitivity tests further reveal that the magnetic island width scales with the square root of the ferritic plate thickness. A substantial magnetic island persists at the S-O location, even when the thickness is reduced to as little as 2 mm. 

\begin{figure}[htbp]
\begin{center}
\includegraphics[clip,width=15.0cm]{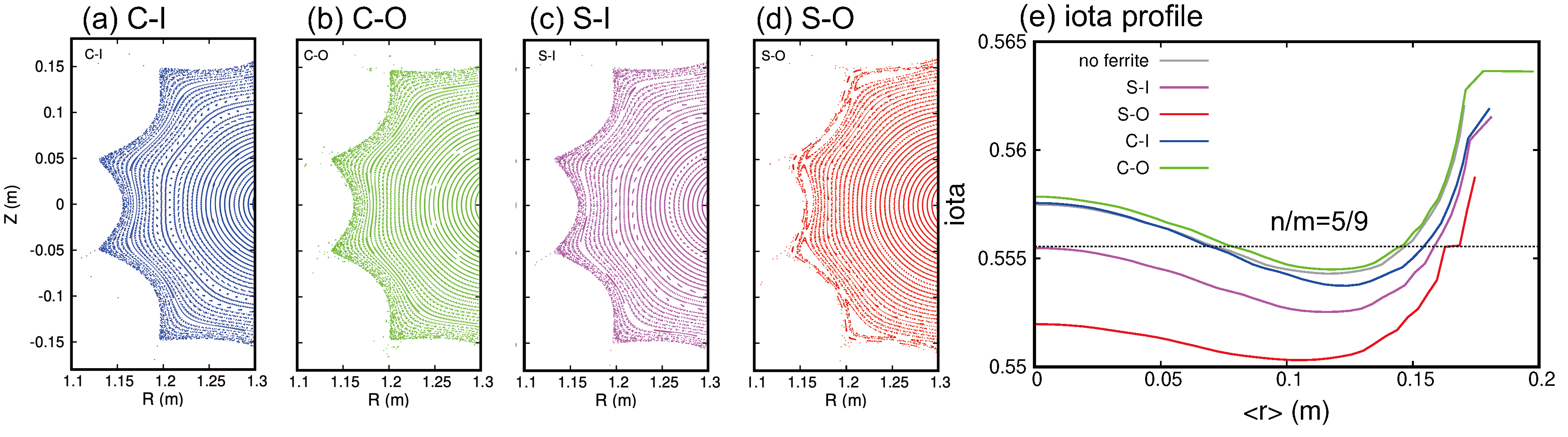}
\caption{Changes in the magnetic configuration at $B_0=1.25$ T when one of the four toroidally periodic ferritic steel plates is vertically displaced by 1.6 cm: (a)--(d) Poincar\'{e} plots of the magnetic field lines and (e) rotational transform profiles.}
\label{fig:ferrite_misalign}
\end{center}
\end{figure}

\begin{figure}[htbp]
\begin{center}
\includegraphics[clip,width=15.0cm]{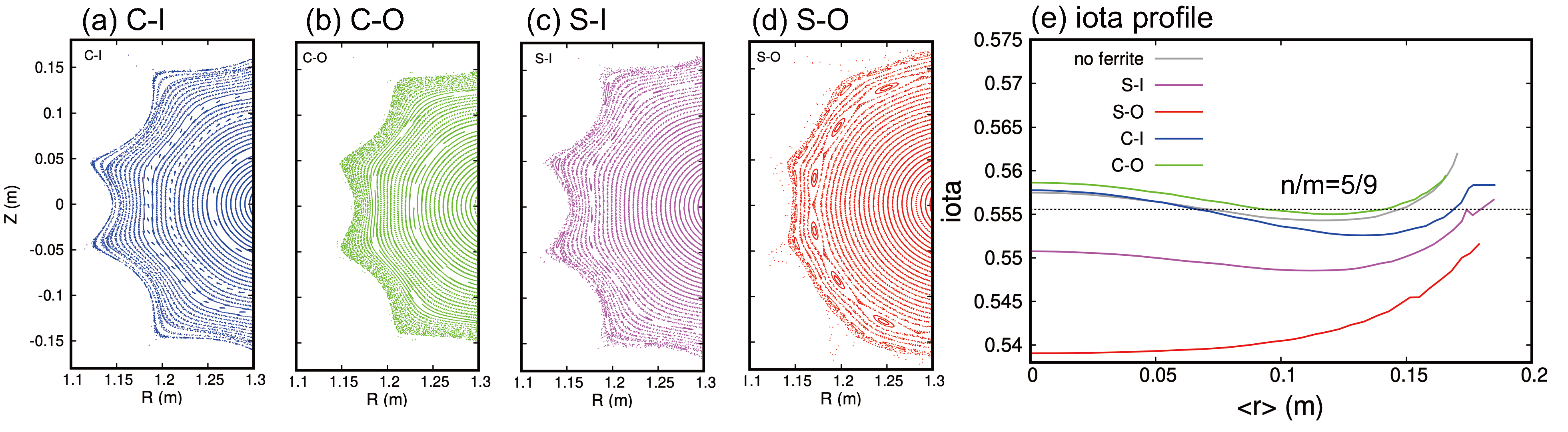}
\caption{Changes in the magnetic configuration at $B_0=0.42$ T when one of the four toroidally periodic ferritic steel plates is vertically displaced by 1.6 cm: (a)--(d) Poincar\'{e} plots of the magnetic field lines and (e) rotational transform profiles.}
\label{fig:ferrite_misalign_third}
\end{center}
\end{figure}

As expected, these results have emphasised that preserving the intrinsic toroidal periodicity is crucial for avoiding large vacuum islands in the presence of ferritic materials. In practice, however, perfect symmetry is difficult to achieve because of installation errors, port access, and engineering constraints. To assess the sensitivity to such asymmetries, figures~\ref{fig:ferrite_misalign} and \ref{fig:ferrite_misalign_third} show the effect of vertically displacing one of the four periodic plates by 1.6 cm at $B_0=1.25$ and 0.42 T, respectively. Consistent with the single-plate results, the configuration is most vulnerable at the S-O location. While the other three locations remain robust even for this centimetre-scale displacement, distortions of the peripheral magnetic surfaces are observed at the S-O location for displacements as small as 4 mm. In particular, when installed at the corner section, no large magnetic islands are observed, even with a weak magnetic field of 0.42 T and the presence of misalignment. These results {indicate} that installation tolerances should be evaluated individually based on the local magnetic sensitivity of each location, rather than applying a uniform standard.

\section{Passive magnetic-configuration modification using ferritic steel plates}\label{sec:application}
\subsection{Passive field modification}
The magnetic field generated by ferritic steel is intrinsically limited by the saturation magnetisation of the material $({\mu_0}M_s\approx 2$ T). Consequently, in reactor-scale devices where the equilibrium magnetic field is much stronger, ferritic perturbations are expected to remain relatively small. However, the installation of ferritic steel plates offers a means of {passive} configuration {modification in} present-day experimental devices, especially when operated at relatively low magnetic fields such that the field produced by the ferritic steel can become comparable to the equilibrium magnetic field. 

Under such conditions, ferritic steel plates can be utilised as passive elements to smooth out local field non-uniformities, analogously to the reduction of the toroidal field ripple in tokamak devices. As discussed above, ferritic materials naturally draw the surrounding magnetic flux toward themselves, also enabling the localised modification of the plasma boundary shape without the need for additional power supplies or complex coil systems. Such passive tailoring provides an interesting experimental method for investigating the effect of localised magnetic elements on three-dimensional configurations. Below we show two numerical examples for Heliotron J.

\subsection{Improvements of {the} high-bumpiness configuration}
\begin{figure}[htbp]
\begin{center}
\includegraphics[clip,width=15.0cm]{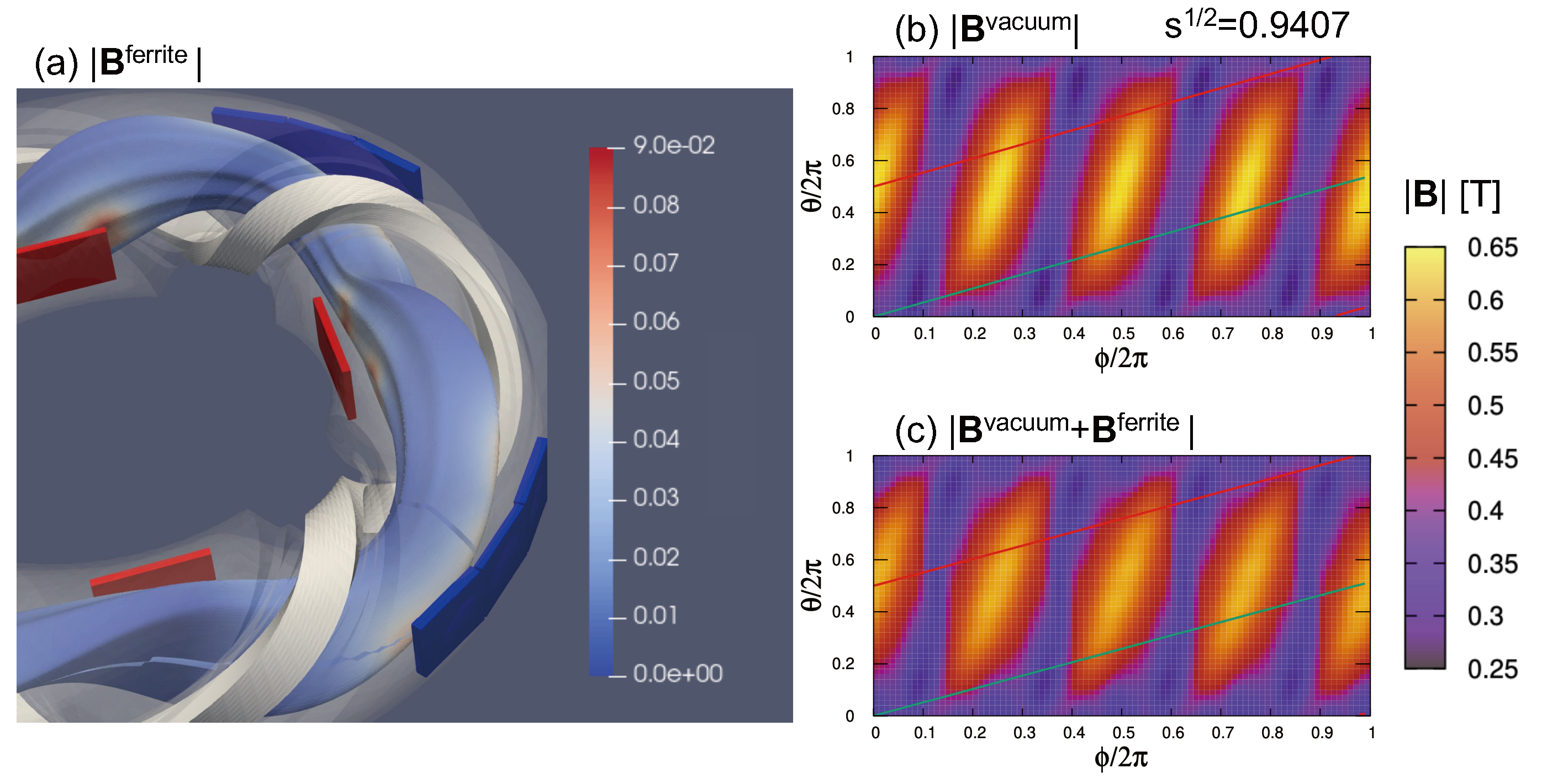}
\caption{{(a) Ferritic steel arrangement and resulting magnetic field strength distribution. (b) Magnetic field strength distribution on the magnetic surface at $\rho=\sqrt{s}=0.9407$ without ferritic steel, where $s$ is the normalised toroidal flux. (c) Corresponding distribution including the ferritic magnetic field}}
\label{fig:optimized}
\end{center}
\end{figure}

\begin{figure}[htbp]
\begin{center}
\includegraphics[clip,width=15.0cm]{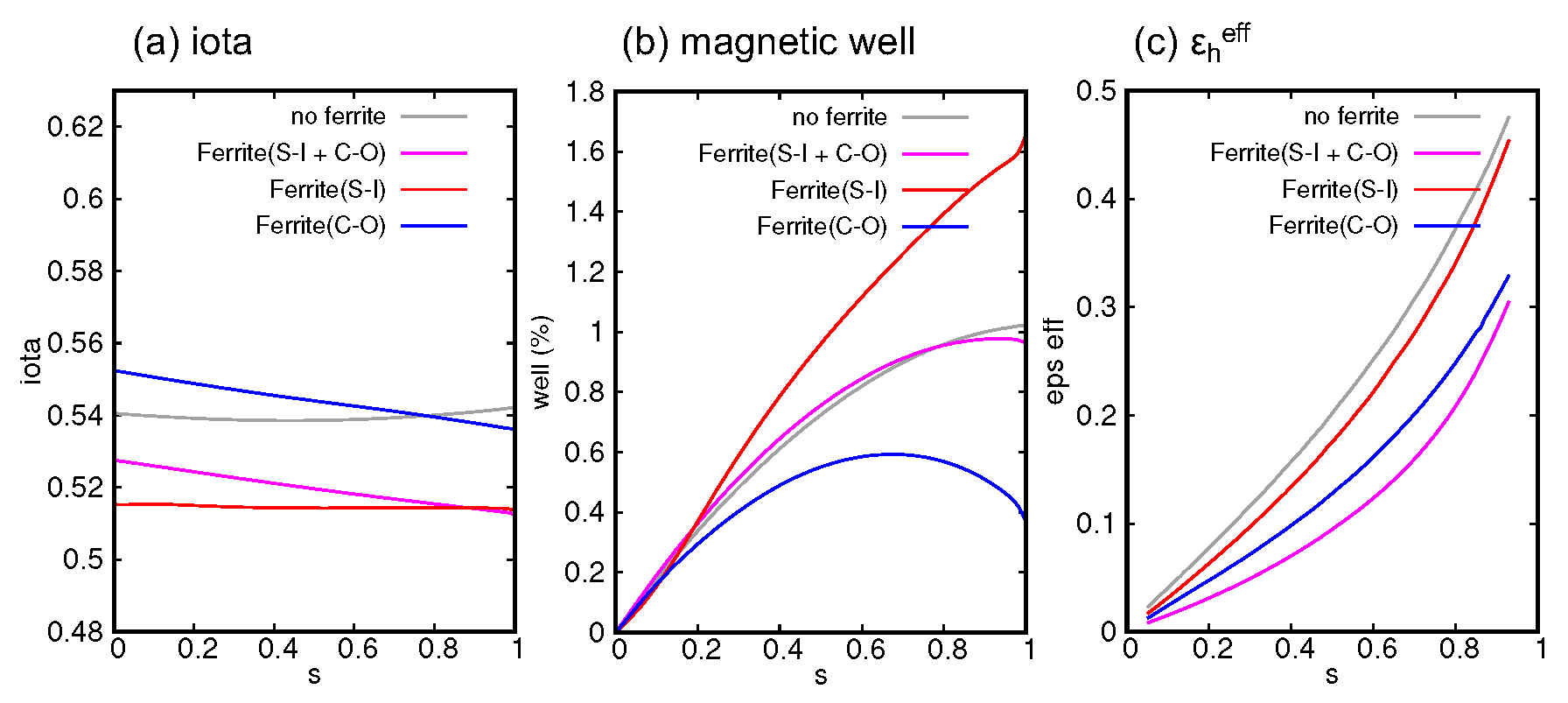}
\caption{Characteristics of the magnetic configuration modified by ferritic steel: (a) rotational transform, (b) magnetic well depth, and (c) effective helical ripple $\epsilon_{\rm eff}^h$.}
\label{fig:optimized_prop}
\end{center}
\end{figure}

Here we consider the high-bumpiness configuration (TA:TB = 5:1) of Heliotron J at $B_0=0.42$ T. In this configuration, the discreteness of the toroidal field coils leaves local ripple structures on the outer side of the corner sections in a similar appearance to the toroidal field ripple in tokamaks. Hence we consider reducing the effective helical ripple $(\epsilon_h^{\rm eff})$ \cite{Nemov1999} by placing ferritic plates on the outer side of the corner sections (C-O; blue) to flatten such a local ripple well. Three ferritic steel plates of a width of 30 cm, height of 24 cm {with} thicknesses of {either} 30 mm {or} 40 mm are placed on the outboard side of the corner sections, as shown in figure~\ref{fig:optimized}(a). The blue curves in figure~\ref{fig:optimized_prop} show the characteristics of the magnetic configuration modified by ferritic steel, showing that this arrangement substantially reduces $\epsilon_h^{\rm eff}$, where $\epsilon_h^{\rm eff}$ was calculated using the {\small NEO} code \cite{Nemov1999}. It is worth noting in figure~\ref{fig:optimized_prop}(b) that this modification simultaneously decreases the depth of the vacuum magnetic well [figure~\ref{fig:optimized_prop}(b)], a side effect consistent with the tendency of the magnetic circuit to draw the flux toward the low-field side. 

To compensate for this reduction, additional plates of a width of 40 cm, height of 24 cm, and thickness of 30 mm are introduced on the high-field side, specifically, the inboard side of the straight section (S-I; red), as shown in figure~\ref{fig:optimized}(a). These plates also weaken the local peak of the magnetic field near the $B_{\rm min}$ wells associated with the toroidicity, contributing to aligning $B_{\rm min}$ well depth {and potentially} improving the confinement of deeply trapped particles. The comparison of the mod-$B$ contour near the plasma surface is displayed in figures~\ref{fig:optimized}(b) and (c). As shown in figure~\ref{fig:optimized_prop}(c), the S-I plates also contribute to the reduction of $\epsilon_h^{\rm eff}$, and their addition further enhances the improvement compared to the C-O only case. By combining these two locations, the configuration restores the vacuum magnetic well to approximately 1$\%$ [figure~\ref{fig:optimized_prop}(b)] while achieving the minimum effective helical ripple. 

The relatively large ferritic-steel thicknesses considered here are intended to explore the range of magnetic modification achievable with passive ferritic elements. In Heliotron J, however, the thickness of in-vessel materials is restricted to approximately 20 mm or less, and direct exposure of the ferritic elements to the discharge should also be avoided. Practical implementation would therefore require further optimisation of the ferritic-steel location, area, and thickness. The external coil currents provide additional degrees of freedom for this purpose \cite{Kin2026}, allowing the background vacuum configuration and ferritic-steel arrangement to be optimised in combination.

\subsection{Boundary shape modification using {ferritic steel}}
\begin{figure}[htbp]
\begin{center}
\includegraphics[clip,width=12.0cm]{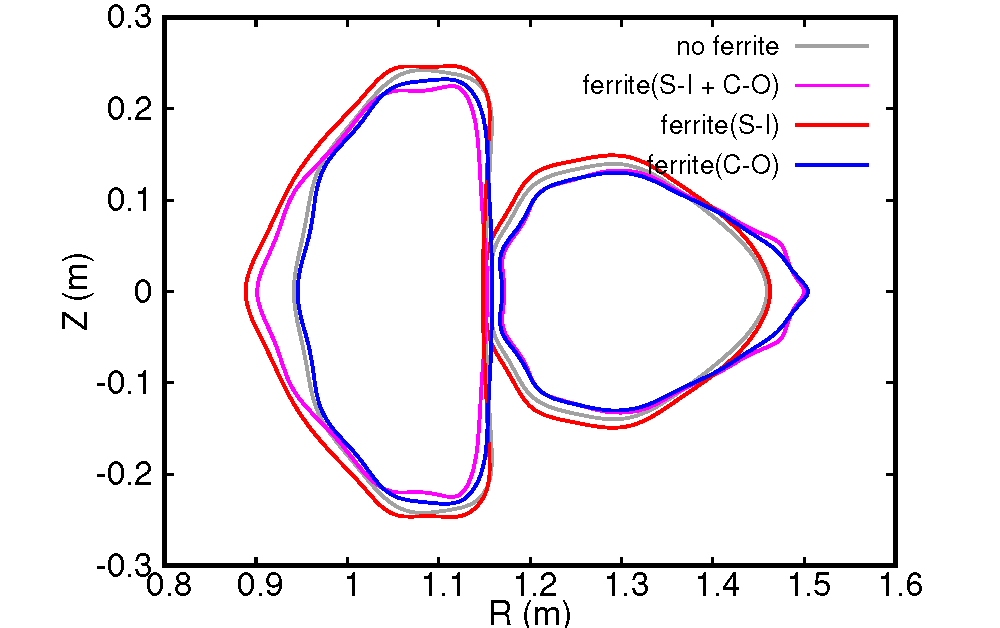}
\caption{Plasma boundary shapes used in the VMEC calculations at the straight (left) and corner (right) sections. The boundaries are chosen slightly inside the corresponding vacuum last closed flux surfaces.}
\label{fig:optimized_boundary}
\end{center}
\end{figure}

Figure~\ref{fig:optimized_boundary} shows the VMEC boundary shape for different ferritic-steel arrangements, corresponding to the ferritic-steel insert cases of figure \ref{fig:optimized}. The unperturbed configuration is shown in grey, while magenta represents ferritic steel installed at both the inboard side of the straight section and the outboard side of the corner section. The red curves correspond to configurations with ferritic steel installed at only one of these locations. It is seen that ferritic steel plates modify not only the magnetic-field-strength distribution and rotational transform but also the geometry of magnetic surfaces in this weak equilibrium field regime {at $B_0=0.42$ T}. Figure~\ref{fig:optimized_boundary} illustrates that the outer magnetic surfaces are displaced toward the ferritic steel plates. This behaviour is consistent with the ferritic material drawing the surrounding magnetic flux toward itself.

An intriguing extension of this concept is to produce magnetic configurations that break the stellarator symmetry \cite{Dewar1998}. By arranging ferritic elements asymmetrically, {a} localised symmetry-breaking {perturbation} can be introduced while keeping the external coil system itself symmetric. An important feature of this approach is not necessarily to generate a large global symmetry breaking, but rather to provide a controllable local perturbation to an otherwise symmetric magnetic configuration. Such configurations offer a unique way to investigate the consequences of symmetry breaking in stellarator/heliotron plasmas, similar to research in tokamaks where up-down asymmetry is known to influence intrinsic rotation and momentum transport \cite{Ball2014,Ball2018}. 

\begin{figure}[htbp]
\begin{center}
\includegraphics[clip,width=10.0cm]{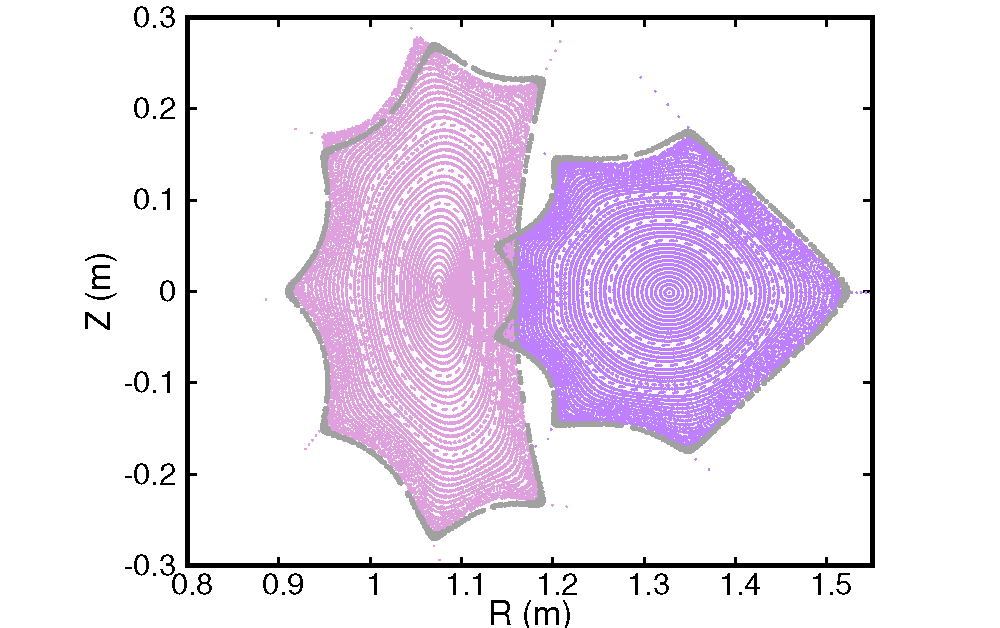}
\caption{Vacuum magnetic surfaces (Poincar\'{e} plots) at the straight (left) and corner (right) sections for a stellarator-asymmetric ferritic-steel arrangement, compared with the corresponding unperturbed surfaces without ferritic steel. Ferritic-steel plates are installed with the $M=4$ toroidal periodicity on the upper-inboard side of the straight sections (top left), thereby breaking the stellarator symmetry while preserving the toroidal periodicity.}
\label{fig:ferrite_asym}
\end{center}
\end{figure}

Figure~\ref{fig:ferrite_asym} illustrates an example in the standard configuration of Heliotron J, where ferritic-steel plates are placed with the M=4 toroidal periodicity on the upper-inboard side of the straight sections. The resulting perturbation remains localised near the plate location and does not adhere to the intrinsic stellarator symmetry of the configuration. This example illustrates the possibility of introducing a localised symmetry-breaking perturbation without modifying the external coil system, although the resulting boundary deformation at $B_0=0.42$ T remains relatively limited.

To obtain a larger asymmetric deformation with a feasible ferritic-steel thickness, the equilibrium magnetic field may be reduced further, e.g., down to $B_0=0.1$ T. Preliminary {\small KFERRITE} calculations indicated this possibility; however, the ferritic steel enters the unsaturated regime under such conditions. The present results also point to the unsaturated magnetisation regime as a possible extension of the parameter space for passive magnetic-field modification in Heliotron J. While this regime may provide additional flexibility beyond the location and thickness of the ferritic elements considered here, its quantitative assessment requires a self-consistent treatment of the ferritic magnetisation.

\section{Discussion and Conclusions}\label{sec:conclusion}
In this study, motivated by the use of RAFM steel in present experimental device{s} and {their potential application in} future fusion reactors, we have numerically investigated the magnetic response of a low-shear heliotron configuration to localised ferritic perturbations. The {\small KFERRITE} code was developed and benchmarked against experimental data from JT-60U. 

The main conclusion is that the magnetic response in a stellarator/heliotron is governed not only by the amplitude of the ferritic magnetic field but also by its spatial coupling to the background magnetic field. While asymmetry only in the poloidal direction is expected for tokamaks \cite{Boozer2011}, the sensitivity is highly non-uniform in both the poloidal and toroidal directions in the stellarator/heliotron device. The outer side of the straight sections (S-O) was identified as the most sensitive location in Heliotron J. At this position, even a weak ferritic perturbation can produce significant changes in the rotational transform and the formation of $(m,n)=(9,5)$ magnetic island chains. {Fourier analysis of the $n=5$ perturbation showed that the S-O location produces a broad poloidal mode spectrum and enhances the resonant $(9,5)$ component, consistent with the observed ordering of the magnetic-island widths.} Numerical sensitivity tests {showed} that the magnetic island width scales with the square root of the ferritic plate thickness, {such that a sizable island persists even for relatively thin ferritic components at the S-O location. While some locations are robust against centimetre-scale displacements, the S-O location is sensitive to millimetre-scale deviations from the intrinsic field periodicity.} 

These results also have implications for the use of ferritic materials in future stellarator/heliotron reactors, where the magnetic field errors associated with the ferritic field can affect the optimum coil configuration \cite{Landreman2026}. {Identifying} and avoiding highly sensitive locations may therefore help preserve the integrity of the magnetic field with minimal modification to the external coil system. This consideration is also related to the stellarator coil optimisation under finite engineering tolerances \cite{Lobsien2018}. A more quantitative analysis of such localised sensitivity across different installation locations, {for example by} extending {approaches such as} the method of \cite{Zhu2019}, remains for future work.

The present analysis considers the vacuum magnetic configuration as a well-defined reference for isolating the geometrical sensitivity to the location of the ferritic perturbation. At finite plasma pressure, changes in the rotational transform profile and local magnetic shear can shift rational surfaces and modify island widths, while plasma response may screen or amplify resonant perturbations. A self-consistent three-dimensional treatment of finite-$\beta$ equilibria containing magnetic islands introduces additional equilibrium and response physics that is distinct from the geometrical effect investigated here. The island widths reported in this study should not be directly extrapolated quantitatively to finite-$\beta$ plasmas. We also note that practical implementation would require additional consideration of experimental constraints not included in the present numerical analysis, including allowable ferritic-steel dimensions, material temperature and cooling requirements, plasma-wall interactions such as arcing, erosion and impurity generation, and compatibility with plasma diagnostics.

Beyond its role as a source of error fields, we have discussed whether ferritic steel can provide a passive means for magnetic configuration tailoring in present-day low-field experiments. Ferritic steel can locally modify the magnetic field through its magnetisation in the background equilibrium field. Numerical examples showed that an appropriate arrangement of ferritic plates can reduce the effective helical ripple while preserving the vacuum magnetic well depth. An asymmetric ferritic-steel arrangement can also introduce a localised perturbation that breaks stellarator symmetry without modifying the external coil system, providing a means to investigate the effect of symmetry breaking on transport properties. Although the deformation obtained in Heliotron J at $B_0=0.42$ T is limited, lower-field operation down to $B_0\sim0.1$ T may provide greater flexibility, provided that the ferritic magnetisation is treated self-consistently. These results indicate that ferritic materials placed at different toroidal and poloidal locations can provide passive degrees of freedom for selectively modifying low-shear heliotron magnetic configurations.

\ack
The authors would like to thank M. Honda, K. Shinohara, K. Tobita, and N. Aiba for providing the JT-60U ferritic magnetic field benchmark data. This work was partially supported by NIFS Fundamental Facility Type Collaboration Research Program (KFFT003), the ZE Research Program, Institute of Advanced Energy, Kyoto University (Reference No. ZE2026B-38), QST Research collaboration for Fusion DEMO, and by Grants-in-Aid for Scientific Research (MEXT KAKENHI Grant No. 25K00982).

% shaping coils: plasma boundary上のb-normalを調整する


\section*{References}
\begin{thebibliography}{99}
\bibitem{Kohyama1996} Kohyama A {\it et al.} 1996 {\it J. Nucl. Mater.} {\bf 233-237} 138
\bibitem{Tanigawa2011} Tanigawa H {\it et al} 2011 {\it J. Nucl. Mater.} {\bf 417} 9
\bibitem{Tanigawa2017} Tanigawa H {\it et al.} 2017 {\it Nucl. Fusion} {\bf 57} 092004 
\bibitem{Tavassoli2014} Tavassoli F {\it et al} 2014 {\it J. Nucl. Mater.} {\bf 455} 269
\bibitem{Jitsukawa2002} Jitsukawa S {\it et al} 2002 {\it J. Nucl. Mater.} {\bf 307-311} 179
\bibitem{Mergia2008} Mergia K and Boukos N 2008 {\it J. Nucl. Mater.} {\bf 373} 1
\bibitem{Gorley2021} Gorley M {\it et al} 2021 {\it Fusion Eng. Des.} {\bf 170} 112513
\bibitem{Giancarli2012} Giancarli G M {\it et al} 2012 {\it Fusion Eng. Des} {\bf 87} 395
%
\bibitem{Turner1978} Turner L R {\it et al} 1978 {\it Iron shielding to decrease toroidal field ripple in a tokamak reactor} Proc. 3rd Topical Meeting on Technology of Controlled Nuclear Fusion (Santa Fe) p 883
\bibitem{Sheffield1993} Sheffield G V 1993 {\it The use of iron shims to reduce the toroidal field ripple in tokamaks} PPPL-2876
\bibitem{Ane1994} Ane J M 1994 {\it Ripple reduction with magnetic inserts and saddle coils} Proc. 18th SOFT (Karlsruhe) p. 723
\bibitem{Tobita2003} Tobita K {\it et al} 2003 {\it Plasma Phys. Control. Fusion} {\bf 45} 133
%
\bibitem{Sato1998} Sato M {\it et al} 1998 {\it J. Nucl. Mater.} {\bf 258-63} 1253 
\bibitem{Kawashima2001} Kawashima H {\it et al} 2001 {\it Nucl. Fusion} {\bf 41} 257
\bibitem{Shinohara2007} Shinohara K {\it et al} 2007 {\it Nucl. Fusion} {\bf 47} 997
\bibitem{Shinohara2003} Shinohara K {\it et al} 2003 {\it Nucl. Fusion} {\bf 43} 586
\bibitem{Yoshida2006} Yoshida M {\it et al} 2006 {\it Plasma Phys. Control. Fusion} {\bf 48} 1673
\bibitem{Honda2014} Honda M {\it et al} 2014 {\it Nucl. Fusion} {\bf 54} 114005 
\bibitem{Tsuzuki2006} Tsuzuki K {\it et al} 2006 {\it Fusion Sci. Technol.} {\bf 49} 197
%\bibitem{Sato2000} Sato M {\it et al} 2000 {\it Fusion Eng. Des.} {\bf 51-52} 1071
%
\bibitem{Harmeyer1999} Harmeyer E {\it et al} 1999 {\it The Effect of Ferritic Structural Material on the Magnetic Field of Stellarators} IPP Report IPP III/241
\bibitem{Ji2017} Ji X {\it et al} 2017 {\it Fusion Eng. Des.} {\bf 125} 631
%
\bibitem{Landreman2026} Landreman M {\it et al} 2026 {\it Efficient calculation of magnetic fields from ferromagnetic materials near strong electromagnets, and application to stellarator coil optimization} arXiv:2511.17305v2 [physics.plasm-ph] 12 Jan 2026
\bibitem{Wakatani2000} Wakatani M {\it et al} 2000 {\it Nucl. Fusion} {\bf 40} 569 
\bibitem{Obiki2001} Obiki T {\it et al} 2001 {\it Nucl. Fusion} {\bf 41} 833 
%
\bibitem{Beidler2021} Beidler C D {\it et al} 2021 {\it Nature} {\bf 596} 221 
\bibitem{Bandyopadhyay2025} Bandyopadhyay I {\it et al} 2025 Nucl. Fusion {\bf 65} 103001
%
\bibitem{Yamazaki1993} Yamazaki K {\it et al} 1993 {\it Fusion Eng. Des.} {\bf 20} 79
\bibitem{Andreeva2009} Andreeva T {\it et al} 2009 {\it Fusion Eng. Des.} {\bf 84} 408
\bibitem{Shoji2023} Shoji M {\it et al} 2023 {\it Plasma Fusion Res.} {\bf 18} 2405026
%
\bibitem{Jaenicke1993} Jaenicke R {\it et al} 1993 {\it Nucl. Fusion} {\bf 33} 687
\bibitem{Morisaki2010} Morisaki T {\it et al} 2010 {\it Fusion Sci. Technol.} {\bf 58} 465
\bibitem{Pedersen2016} Sunn Pedersen T {\it et al} 2016 {\it Nat. Commun.} {\bf 7} 13493 
%
\bibitem{Lazerson2018} Lazerson S A {\it et al} 2018 {\it Plasma Phys. Control. Fusion} {\bf 60} 124002
%
\bibitem{Boozer2011} Boozer A H 2011 {\it Fusion Sci. Technol.} {\bf 59} 561 
\bibitem{Pharr2024} Pharr M {\it et al} 2024 {\it Nucl. Fusion} {\bf 64} 126025
\bibitem{Zhu2019} Zhu Caoxiang {\it et al} 2019 {\it Nucl. Fusion} {\bf 59} 126007 
\bibitem{Cao2026} Cao Yuhao {\it et al} 2026 Eur. Phys. J. D {\bf 80} 22
%
%
\bibitem{Helander2020} Helander P {\it et al} 2020 {\it Phys. Rev. Lett.} {\bf 124} 095001
\bibitem{Hammond2020} Hammond K C {\it et al} 2020 {\it Nucl. Fusion} {\bf 60} 106010
\bibitem{Qian2022} Qian T {\it et al} 2022 {\it Nucl. Fusion} {\bf 62} 084001
\bibitem{Gates2025} Gates D A {\it et al} 2025 {\it Nucl. Fusion} {\bf 65} 026052
\bibitem{Kruger2025} Kruger T G {\it et al} 2025 {\it Nucl. Fusion} {\bf 65} 026051
\bibitem{Kaptangolu2025} Kaptangolu A A 2025 {\it Phys. Rev. E} {\bf 111} 065202
\bibitem{Ku2009} Ku L P and Boozer A H 2009 {\it Phys. Plasmas} {\bf 16} 082506
\bibitem{Elder2024} T. Elder and A. H. Boozer 2024 {\it Phys. Plasmas} {\bf 31} 102501 
\bibitem{Boozer2024} Boozer A H 2024 {\it Phys. Plasmas} {\bf 31} 122505
%
\bibitem{Goodman2023} Goodman A G {\it et al} 2023 {\it J. Plasma Phys.} {\bf 89} 2
\bibitem{Mynick1982} Mynick H E {\it et al} 1982 {\it Phys. Rev. Lett.} {\bf 48} 322
\bibitem{Yokoyama2000} Yokoyama M {\it et al} 2000 {\it Nucl. Fusion} {\bf 40} 161
\bibitem{Nakamura1992} Nakamura Y {\it et al} 1992 {\it J. Plasma Fusion Res. } {\bf 69} 41 
\bibitem{Todoroki1987} Todoroki J 1987 Kakuyugo-Kenkyu {\bf 57} 318 (in Japanese)
%
\bibitem{Hirshman1983} Hirshman S P and Whitson J C 1983 {\it Phys. Fluids} {\bf 26} 3553
\bibitem{Sanchez2000} Sanchez R {\it et al} 2000 {\it Plasma Phys. Control. Fusion} {\bf 42} 641
%
\bibitem{Urata2003} Urata K 2003 {\it "Development of FEMAG: Calculation Code of Magnetic Field Generated by Ferritic Plates in the Tokamak Devices} JAERI-Data/Code 2003-005  
\bibitem{Shiba1997} Shiba K {\it et al} 1997 {\it Properties of Low Activation Ferritic Steel F82H IEA Heat - Interim Report of IEA Round-robin Tests (1) -} JAERI-Tech 97-038 (in Japanese) 
\bibitem{Nakayama1999} Nakayama T {\it et al} 1999 {\it J. Nucl. Mater.} {\bf 271-272} 491
\bibitem{Aharoni1998} Aharoni A 1998 {\it J. Appl. Phys.} {\bf 83} 3432 
%
\bibitem{Lobsien2018} Lobsien Jim-Felix {\it et al} 2018 {\it Nucl. Fusion} {\bf 58} 106013 
%
\bibitem{Nemov1999} Nemov V V {\it et al} 1999 {\it Phys. Plasmas} {\bf 6} 4622
\bibitem{Kin2026} {Kin F {\it et al} 2026 {\it Plasma Phys. Control. Fusion} {\bf 68} 085050}
%
\bibitem{Dewar1998} Dewar R L and Hudson S R 1998 {\it Physica D} {\bf 112} 275
\bibitem{Ball2014} Ball J {\it et al} 2014 {\it Plasma Phys. Control. Fusion} {\bf 56} 095014 
\bibitem{Ball2018} Ball J {\it et al} 2018 {\it Nucl. Fusion} {\bf 58} 026003
%
\end{thebibliography}
\end{document}